\documentclass[aps, prx, reprint, superscriptaddress, longbibliography, floatfix]{revtex4-2}
\usepackage{graphicx}
\usepackage{placeins}
\usepackage{dcolumn}
\usepackage{bm}
\usepackage{physics}
\DeclareMathAlphabet{\pazocal}{OMS}{zplm}{m}{n}
\usepackage{amsmath}
\usepackage{amsfonts}
\usepackage{mathrsfs}
\usepackage{subfigure}
\usepackage{color,soul}
\usepackage{xcolor}
\usepackage[colorlinks=true,linkcolor=blue,anchorcolor=red,citecolor=blue, urlcolor=blue]{hyperref}

\begin{document}

\title{Ferromagnetic ordering in mazelike stripe liquid of a dipolar six-state clock model}

\author{Quentin Simmons}
\affiliation{Department of Physics, University of Virginia, Charlottesville, VA 22904, USA}

\author{Shi-Zeng Lin}
\affiliation{Center for Nonlinear Studies and Theoretical Division, Los Alamos National Laboratory, Los Alamos, NM 87545, USA}

\author{Gia-Wei Chern}
\affiliation{Department of Physics, University of Virginia, Charlottesville, VA 22904, USA}

\begin{abstract}
We present a comprehensive numerical study of a six-state clock model with a long-range dipolar type interaction. This model is motivated by the ferroelectric orders in the multiferroic hexagonal manganites. At low temperatures, trimerization of local atomic structures leads to six distinct but energetically degenerate structural distortion, which can be modeled by a six-state clock model. Moreover, the atomic displacements in the trimerized state further produce a local electric polarization whose sign depends on whether the clock variable is even or odd. These induced electric dipoles, which can be modeled by emergent Ising degrees of freedom, interact with each other via long-range dipolar interactions. Extensive Monte Carlo simulations are carried out to investigate low temperature phases resulting from the competing interactions.  Upon lowering  temperature, the system undergoes two Berezinskii-Kosterlitz-Thouless (BKT) transitions, characteristic of the standard six-state clock model in two dimensions. The dipolar interaction between emergent Ising spins induces a first-order transition into a ground state characterized by a three-fold degenerate stripe order. The intermediate phase between the discontinuous and the second BKT transition corresponds to a maze-like hexagonal liquid with short-range stripe ordering. Moreover, this intermediate phase also exhibits an unusual ferromagnetic order with two adjacent clock variables occupying the two types of stripes of the labyrinthine pattern. 
\end{abstract}
\date{\today}
\maketitle

\section{Introduction}

\label{sec:intro}

Systems with long-range interactions are sources of complex emergent orders and unusual phase transitions~\cite{Dauxois02,Mukamel08}. Perhaps one of the simplest long-range interacting systems is the two-dimensional (2D) ferromagnetic Ising model with dipolar interaction, which can be realized in ultra-thin magnetic films~\cite{DeBell2000}. Despite its simplicity, the 2D dipolar Ising ferromagnet exhibits a rich phase diagram~\cite{Booth1995,MacIsaac1995,Toloza98,Rastelli06,Cannas2006,Pighin07,Rastelli2007,Vindigni08,Rizzi10,Fonseca12,Horowitz15,Bab16,Ruger2012,Komatsu18,Bab19,Jin22}. The short-range ferromagnetic interaction favors a single domain phase while the dipolar interaction tends to stabilize a multi-domain structure. The competition between these two interactions gives rise to a plethora of phases which have not yet been fully understood~\cite{Vaterlaus2000,Portmann2003,Wu2004,Cannas2004,Portmann2006}. It has been established that an arbitrary small dipolar interaction stabilizes a stripe domain pattern, with long-range orientational order and quasi-long-range positional order~\cite{Booth1995, MacIsaac1995}. Upon increasing temperature, the system undergoes an order-disorder transition into a liquid phase with strong short-range correlations. The melting of the stripe order is characterized by the breaking of discrete orientational symmetry and is accompanied by a sharp peak in the specific heat.

This correlated liquid phase, called the tetragonal (hexagonal) liquid in the square (honeycomb) lattice model, crosses over into the uncorrelated paramagnetic phase at higher temperatures. The intermediate liquid phase is characterized by well-developed ferromagnetic domains that form maze-like patterns~\cite{Booth1995,Ruger12}. The short-range stripe-like correlation as well as the discrete lattice symmetry are preserved in this intermediate phase. Recently an intermediate nematic phase exhibiting orientational order but without positional order was observed in Monte Carlo simulations~\cite{Cannas2006}, which is consistent with a theoretical prediction based on the continuum approximation~\cite{Abanov1995}.  The nature of phase transitions among the various phases and stripe orders is still under debate.

The effects of long-range interactions have also been investigated in other spin models with more complicated internal symmetries~\cite{Maleev76,Prakash90,Maier04,Tomita09,Baek11,Giachetti22,Giachetti2023}. In particular, although 2D systems with continuous degrees of freedom, such as XY or Heisenberg spins, cannot exhibit long-range order at finite temperatures, the presence of dipolar interactions is shown to induce an ordered state that breaks the continuous symmetry~\cite{Maleev76}. The anisotropic nature of the dipolar interaction also introduces a coupling between lattice symmetry and the internal symmetry of spins~\cite{Prakash90}. As a result, the continuous symmetry is effectively reduced to a discrete one depending on the lattice geometry. 

\begin{figure}[t]
\includegraphics[width=0.99\columnwidth]{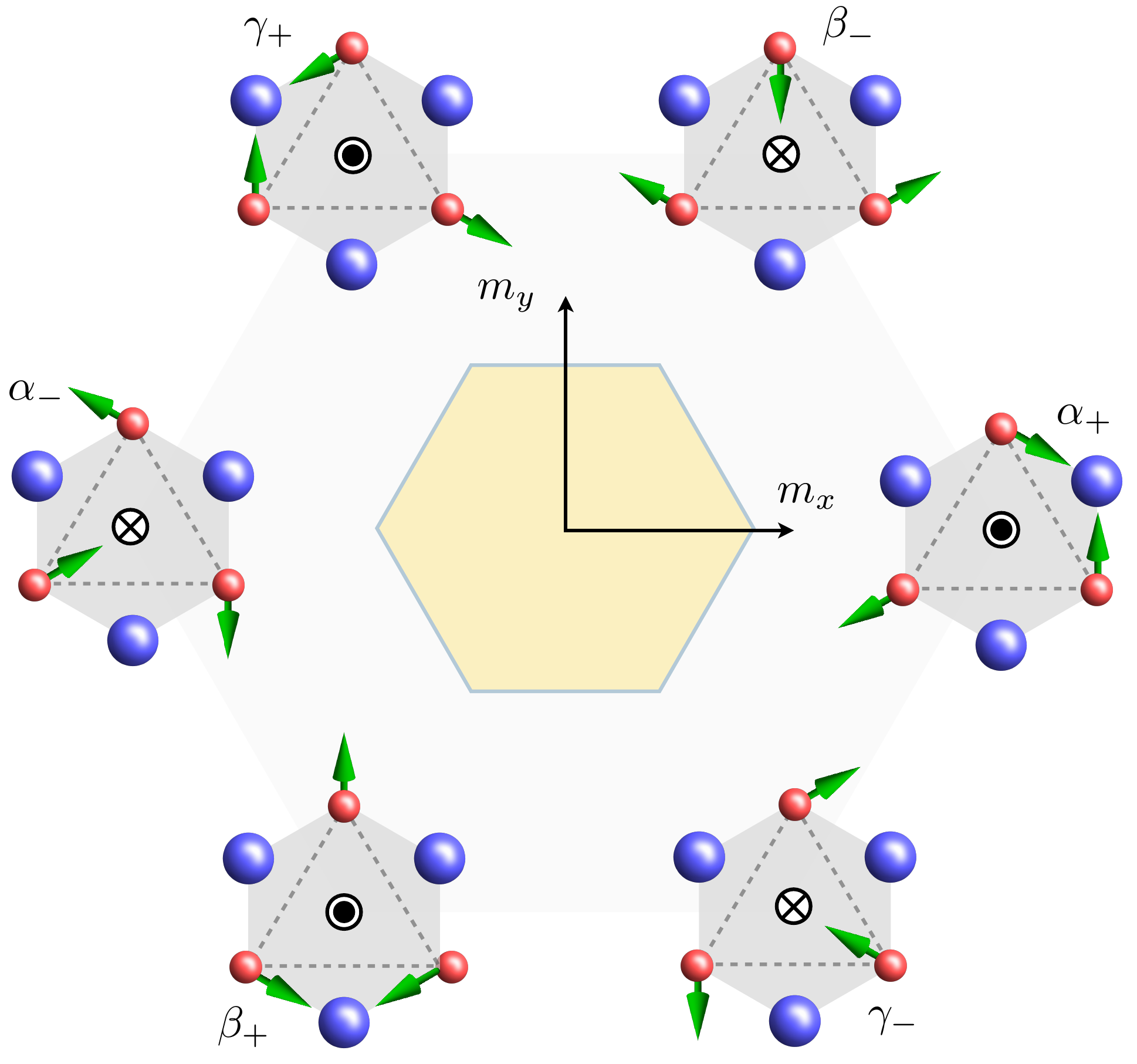}
\caption{\label{fig:trimerization} Polarization scheme of trimerized materials. There are three positive polarization states ($\alpha+, \beta+, \gamma+$) and three negative polarization states ($\alpha-, \beta-, \gamma-$). The symbols $\odot$ and $\otimes$ denote induced electric polarizations $\mathbf P = + P_0 \hat{\mathbf z}$ and $\mathbf P = - P_0 \hat{\mathbf z}$, respectively, that are perpendicular to the triangular layer.   }
\end{figure}

In this paper, we consider an unusual dipolar system characterized by a short-range interaction between discrete spins and a long-range dipolar interaction between emergent Ising variables. This model is relevant for the multiferroic hexagonal manganites, such as $\mathrm{RMnO_3}$ (R denotes rare earth elements). The onset of ferroelectricity in hexagonal manganites is triggered by a structural instability called trimerization\cite{Van2004,Choi2010, Mostovoy2010}, which breaks a three-fold lattice symmetry; see FIG.~\ref{fig:trimerization}. The trimerization is further accompanied by an ionic displacement that produces a net electric dipole moment $\mathbf P$ associated with the structural unit cell. As the polarization can be either parallel or antiparallel to the $c$-axis of the system, it effectively represents an emergent Ising degree of freedom. The system thus exhibits a $Z_3\times Z_2$ symmetry and the short-range interaction between local lattice distortions can be effectively described by a six-state clock model~\cite{szlin2014}.  

While the six-state clock model on a triangular lattice is shown to successfully describe the ferroelectric transition in hexagonal manganites~\cite{szlin2014}, the effects due to the distortion-induced electric polarization remain unexplored. As discussed above, the long-range nature of the dipolar interaction between these local electric moments could introduce additional phases at low temperatures. Moreover, due to the $Z_2$ nature of the induced electric dipole moment $\mathbf P = \pm P_0 \,\hat{\mathbf z}$, where $P_0$ is the amplitude of the induced electrical polarization, dipolar interaction between the effective Ising degrees of freedom is expected to stabilize stripe orders.

Here we consider a six-state clock model with dipolar interaction between emergent Ising spins as a minimum model for hexagonal manganites. Extensive Monte Carlo simulations are carried out to investigate the thermodynamic behaviors and various low-temperature phases of this model. In the absence of the dipolar interaction, it is well known that the system undergoes two Berezinskii-Kosterlitz-Thouless (BKT) phase transitions associated with breaking of $U(1)$ symmetry, \cite{Jose1977, Challa1986} and a uniform domain is stabilized at low temperatures. With the dipolar interaction, the low temperature single domain is expected to be replaced by striped domains with alternating polarization possessing long-range orientational order and quasi-long-range positional order. Little is known about phases in the intermediate temperature region. Here we reveal a novel inhomogeneous phase with broken $U(1)$ symmetry but without any orientational or positional order in the intermediate temperature region. Our prediction can be verified in experiments with $\mathrm{RMnO_3}$ using imaging methods.

In Section \ref{sec:model} we lay out the model Hamiltonian of the six-state Potts model with a dipolar interaction, as well as the Monte Carlo methods we use to numerically simulate the system. In Section \ref{sec:results} we discuss the thermodynamic evolution of the system with respect to two regimes: a weak dipolar interaction, and a dipolar interaction comparable to the ferromagnetic interaction. Finally, in Section \ref{sec:summary} we present a conclusion and discuss outlook of our results.

\section{Model and Methods}

\label{sec:model}

\subsection{Six-state clock model with dipolar Ising interaction}

Motivated by recent experimental results on hexagonal manganites RMnO$_3$, notable for their multiferroic properties, here we consider a generalized six-state clock model incorporating the long-range dipolar interactions on a 2D triangular lattice. Ferroelectricity in RMnO$_3$ is a by-product of trimerization of the material, where a lattice distortion caused by the atomic radii mismatch between R and Mn triples the size of the unit cell. The three distinct trimerizatons are denoted as $\alpha$, $\beta$, and $\gamma$. As discussed above, the trimerization is followed by a subsequent ionic displacement that breaks an additional $Z_2$ symmetry (sign of the electric polarization along the $c$ axis, perpendicular to the hexagonal plane). The polarization state can then be represented by an Ising variable $\sigma_i = \pm 1$. The lattice distortion associated with a structural unit, represented by site-$i$ on a triangular lattice, can be conveniently labeled by a phase 
\begin{eqnarray}
	\phi_i = \frac{p_i \pi}{3}, \qquad (p_i = 0, 1, 2, \cdots, 5),
\end{eqnarray}
For convenience, the integer $p_i$ will also be referred to as a Potts variable in the following discussion. The 6 different Potts states correspond to the six distinct lattice distortions $(\alpha_+, \beta_-, \gamma_+, \alpha_-, \beta_+, \gamma_-)$; see FIG.~\ref{fig:trimerization}. The cyclic arrangements of these six structural antiphases are determined from the low-energy vortex and anti-vortex configurations of the manganites. In terms of clock angles, the Ising spins are given by
\begin{eqnarray}
	\label{eq:sigma_i}
	\sigma_i = \cos(3 \phi_i).
\end{eqnarray}
A minimum model is given by the following Hamiltonian
\begin{eqnarray}
	\label{eq:H}
	\mathcal{H} = -J \sum_{\langle ij \rangle} \cos\left(\phi_i - \phi_j \right) + \frac{D}{2}\sum_{i, j} \frac{\sigma_i \sigma_j}{r_{ij}^3}
\end{eqnarray}
The first term with $J > 0$ is the nearest-neighbor ferromagnetic interaction between the clock or phase variables, physically corresponding to the alignment of structural distortions in the ground state. The second term with $D > 0$ arises from the long-range electric dipole-dipole interaction $E_{\rm dp} \propto [ \mathbf P_i \cdot \mathbf P_j - 3 (\mathbf P_i \cdot \hat{\mathbf r}_{ij} ) (\mathbf P_j \cdot \hat{\mathbf r}_{ij}) ] / r_{ij}^3$, where $\mathbf r_{ij} = \mathbf r_j - \mathbf r_i$, $\mathbf P_i = \sigma_i P_0 \hat{\mathbf z}$. Note that the model is best regarded as a thin-film realization of some 3D systems, and we use the 3D dipolar interaction for the Ising variables.

It should be noted that the first term in the Hamiltonian~(\ref{eq:H}) describes an effective short-range interaction that produces the six-fold degenerate structural distortions observed in hexagonal manganites. On general grounds, the breaking of a $Z_6$ symmetry can proceed in a number of scenarios~\cite{Cardy80}. In the first scenario, there could be a single discontinuous transition separating the paramagnetic state and the fully ordered state. The system could also undergo two BKT transitions with an intermediate critical phase. Finally, the six-fold symmetry can also be broken in two stages: the system enters a partially ordered phase via an Ising transition (breaking of the $Z_2$ symmetry) and undergoes a second transition of the 3-state Potts universality class (breaking of the $Z_3$ symmetry), or a similar scenario but with the Ising and Potts transitions exchanged. Previous works have shown that the $Z_6$ symmetry in the 2D six-state clock model, or XY model with six-state clock anisotropy, is broken via the two BKT transition scenario. It is intriguing to see whether this ordering scenario is affected by the dipolar term, especially considering the fact that the $Z_2$ part of the six-fold symmetry is associated with a long-range interaction.

On the other hand, as discussed in Sec.~\ref{sec:intro}, the dipolar term favors a maze-like hexagonal liquid phase and a stripe order (in terms of the Ising spins) at low temperatures. While minimization of the short-range $J$ term requires the same clock state for all spins in a given domain or stripe, the long-range clock ordering is disrupted by the multiple domains of opposite Ising spins coexisting in either the hexagonal liquid or stripe ordered states. The short-range ferromagnetic interaction, however, does impose a constraint that the clock states, say $\phi$ and $\phi'$, of two neighboring domains or stripes are next to each other on the clock face, such that $\cos(\phi - \phi') = 1/2$. These considerations lead to unusual disordered yet short-ranged correlated clock configurations associated with the maze-like or stripe patterns.

\subsection{Monte Carlo simulations}

Here we employ classical Monte Carlo (MC) simulations to investigate the intriguing possibilities of low temperature phases. We model our system on an $L \times L$ triangular lattice with periodic boundary conditions; the number of spins is $N = L^2$. Due to the frustrated interactions of the Hamiltonian, well known cluster algorithms cannot be applied to our case. We resort to single spin update with the standard Metropolis-Hastings algorithm. Specifically, for a given site-$i$, the corresponding spin is updated from $\phi_i$ to a random new clock state $\phi'_i$. The acceptance probability of this local spin update is given by $p_{\rm acc} = \min\{ 1, \exp(-\beta \Delta E ) \}$, where $\beta = 1/T$ is the inverse temperature and $\Delta E$ is the change in energy induced by the spin-update. The contribution of the short-range ferromagnetic term to $\Delta E$ can be efficiently computed by considering the update to the interaction energy with the six nearest neighbors. On the other hand, if the update results in a flipped Ising spin $\sigma_i \to -\sigma_i$, the calculation of the dipolar contribution needs to account for interactions with every other spins in the lattice. The resultant $\mathcal{O}(N)$ scaling time-complexity is one fundamental difficulty for Monte Carlo simulations of systems with long-range interactions. 

One sweep of the system corresponds to applying the single-spin update procedure to every lattice site once. The computation complexity of one sweep, a fundamental time unit for MC simulations, thus scales as $\mathcal{O}(N^2)$. Thermodynamic variables were determined by averaging over many sweeps of the system. Depending on $L$, we ran several hundred to several thousand sweeps of the system in order to reach thermal equilibrium before beginning to collect data. We then collected 100,000 - 200,000 data points to determine thermodynamic quantities.

The use of periodic boundary conditions (PBC), while helping reduce finite size effects, also introduces additional difficulty for the calculation of the dipolar energy. Practically, one considers an expanded system, by tiling an enlarged lattice with identical copies of the original $L \times L$ system. The effective interaction between two spins $\sigma_i$ and $\sigma_j$ thus include dipolar interactions of $\sigma_i$ and copies of $\sigma_j$ in all the replicas. The dipolar term can thus be expressed as 
\begin{eqnarray}
	\mathcal{H}_D = \frac{D}{2} \sum_{ij} \sum_{\mathbf R} \frac{\sigma_i \sigma_j}{\left| \mathbf r_i -\mathbf R - \mathbf r_j \right|^3} \equiv \sum_{ij} \mathcal{D}_{ij} \sigma_i \sigma_j,
\end{eqnarray}
where $\mathbf R$ denotes the position of the replicated lattice relative to the central parent system. Here we have also introduced an effective $N\times N$ interaction matrix $\mathcal{D}_{ij}$ computed from the summation over replicas. Importantly, for a given size of the enlarged lattice, this interaction matrix only needs to be calculated once in advance, thus saving significant computation time. 

In our implementations, we have included $400^2$ replicas in the calculation of the effective dipolar interaction matrix $\mathcal{D}$. Instead of the more sophisticated Ewald summation~\cite{Ewald21}, we adopted a direct summation method as discussed in Ref.~\cite{Ruger12}. Special care has also been taken to ensure the discrete lattice symmetry is preserved with the periodic boundary conditions.

\begin{figure}[hbt!]
\includegraphics[width=0.9\columnwidth]{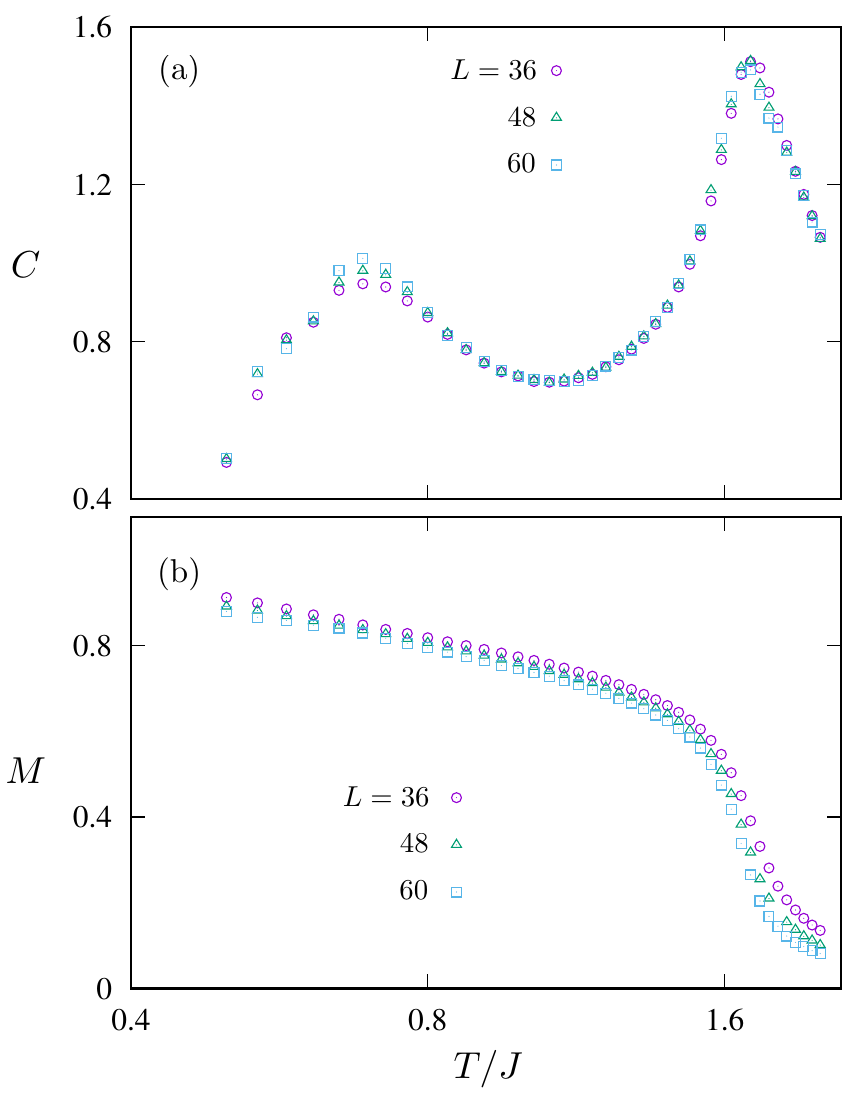}
\caption{\label{f2} (a) Specific Heat $C$ and (b) ferromagnetic order parameter $M$ as a function of $T/J$ calculated from Monte Carlo simulations for a system with $D/J = 0.025$.  In this limit of weak dipolar interaction, the overall thermodynamic behaviors are consistent with the two BKT transitions scenario of a standard 6-state clock model without dipolar interaction.}
\end{figure}

\section{Results}

\label{sec:results}


\begin{figure}[t]
\includegraphics[width=0.99\columnwidth]{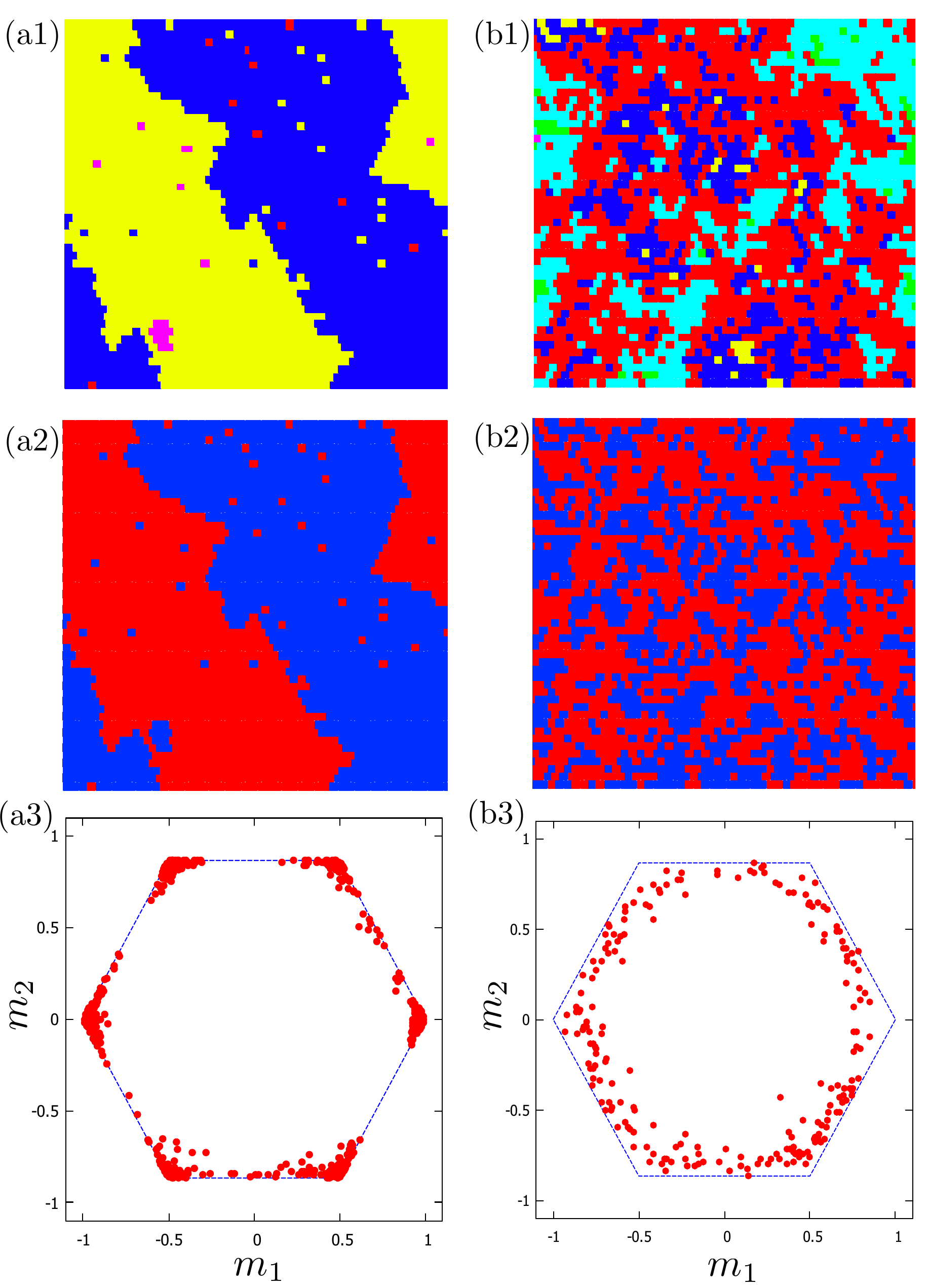}
\caption{\label{f3} (Top) potts configurations, (middle) Ising configurations, and (bottom) histograms of local ferromagnetic order parameter $\mathbf m = (m_1, m_2)$ for (a) the ferromagnetic ordered state at $T < T_2$ and (b) the critical phase for $T_2 < T < T_1$. As $T \rightarrow 0$, (a) will become homogenous and all points in (a3) will be exactly on the corners of the hexagon. Note the emergent O(2) symmetry in (b3) of the critical XY phase. }
\end{figure}

We first analyze the system with a weak dipolar interaction $D \ll J$. The thermodynamic behaviors of the system in this limit are expected to be similar to those of the standard 6-state clock model on a triangular lattice. FIG.~\ref{f2} shows the thermodynamic evolution of the system with $D/J = 0.025$ for 3 different lattice sizes, $L = 36, 48, $ and $60$. We consider the temperature dependence of specific heat and the ferromagnetic order parameter. The specific heat is defined as
\begin{eqnarray}\label{SpecificHeatEq}
	C = (\langle \mathcal{H}^2 \rangle - \langle \mathcal{H} \rangle^2)/NT^2, 
\end{eqnarray}
Where $\langle \cdots \rangle$ means ensemble average through MC sampling. The specific heat, shown in FIG.~\ref{f2}, shows two peaks at $T_1 \approx 1.7375$ and $T_2 \approx 0.6875$. Both the location and height of both specific heat peaks show only a slight dependence on the system sizes. Such weak finite-size scaling effects are characteristic of BKT transitions. Next we consider the ferromagnetic order parameter defined as
\begin{eqnarray}
	 M = \Bigl\langle \Bigl| \frac{1}{N}  \sum_i \mathbf s_i  \Bigr| \Bigr\rangle
\end{eqnarray}
where $\mathbf s_i = (\cos\phi_i, \sin\phi_i)$ is the discrete XY spin at site-$i$. The temperature dependence of $M $ is shown in FIG.~\ref{f2}(b) for three different system sizes. Upon lowering temperature, a pronounced rise of $M$ occurs at the first critical temperature. After which the order parameter seemingly increases linearly towards its maximum $M_{\rm max} = 1$ with decreasing temperature. There is almost no noticeable changes in $M$ when crossing the second critical point $T_2$. Very weak finite size dependence is also observed for the thermodynamic evolution of the ferromagnetic order parameter. Both features are again consistent with the two BKT transitions scenario, as observed in the classical work Ref.~\cite{Challa1986} on the standard 6-state clock model on a square lattice.

The scenario of two BKT transitions was first suggested by theoretical analysis based on a renormalization group argument and low-temperature expansion~\cite{Jose1977}, and was later confirmed by extensive MC simulations and finite size scaling analysis in the classical work of Ref.~\cite{Challa1986}. The two transitions at $T_1$ and $T_2$ are both of the BKT type with an exponentially divergent susceptibility $\chi \sim \xi^{2-\eta}$, where $\xi \sim \exp( a \sqrt{|T - T_{1, 2}|})$ is the temperature-dependent correlation length and $a$ is a non-universal constant. The susceptibility remains infinite in the critical phase in between $T_1$ and $T_2$, while the order parameter exhibits a power-law dependence on system size $M \sim 1/L^\eta$, with the exponent varying continuous from $\eta_1 = 4/9$ at $T_1$ to $\eta_2 = 1/4$ at  $T_2$. 

Snapshots of the real-space Potts and Ising variables sampled from MC simulations for the ferromagnetic phase $(T < T_2)$ and the intermediate critical phase $(T_2 < T < T_1)$ are shown in FIG.~\ref{f3}. The ordered states are characterized by large ferromagnetic domains in both Potts and Ising variables. On the other hand, the critical states exhibit ferromagnetic domains of various sizes in a seemingly disordered fashion. As the correlation between Potts variables decays algebraically in this regime, $\langle \cos(\phi_i - \phi_j) \rangle \sim 1 / r_{ij}^\eta$, the system is expected to show a fractal-like structures. Also importantly, an emergent rotation symmetry is predicted for the XY-like critical phase. To demonstrate this, we define a local block-averaged ferromagnetic order parameter
\begin{eqnarray}
	\mathbf m(\mathbf r) = \frac{1}{N_b} \sum_{i \in B(\mathbf r)} \mathbf s_i,
\end{eqnarray}
where $B(\mathbf r)$ denotes a block of spins centered at $\mathbf r$, and $N_b$ is the number of spins in the block. The domain of this vector order parameter $\mathbf m = (m_1, m_2)$ is a hexagon with the six corners corresponding to the perfectly ordered states.  Histograms of this local order parameter $\mathbf m$ sampled from the ordered and critical phases are shown in FIG.~\ref{f3}(a3) and (b3), respectively. As expected, the order parameter clusters around the six corners of the hexagonal domain. The distribution of $\mathbf m$ in the critical phase, on the other hand, exhibits a circular symmetry characteristic of XY spin systems. This indicates that the 6-state clock anisotropy is an irrelevant perturbation in the intermediate critical phase. 

As noted in Sec.~\ref{sec:intro}, any finite dipolar interaction will stabilize a stripe phase at low enough temperatures. Since the width of the stripes in the ground state increases with decreasing dipolar interaction, our characterization of a ferromagnetic state at $T < T_2$ is subject to the finite size effects. Even the largest system size $L = 60$ might be shorter than the equilibrium stripe width of the system with a small $D = 0.025J$. The study of exact behavior of such small $D$ systems in the thermodynamic limit is beyond our numerical method. On the other hand, the competition between the short-range ferromagnetic ordering and the long-range dipolar interaction is likely to result in meta-stable nonequilibrium state. For example, consider a thermal quench scenario where the temperature is quickly decreased to below $T_2$. Initially multiple ferromagnetic domains of small size are nucleated after the quench. As the typical domain sizes $\lambda$ increase with time, the dipolar interaction becomes more prominent. The coarsening of ferromagnetic domains is likely preempted when $\lambda$ reaches the energetically favored stripe width. 


\begin{figure}[t]
\includegraphics[width=0.9\columnwidth]{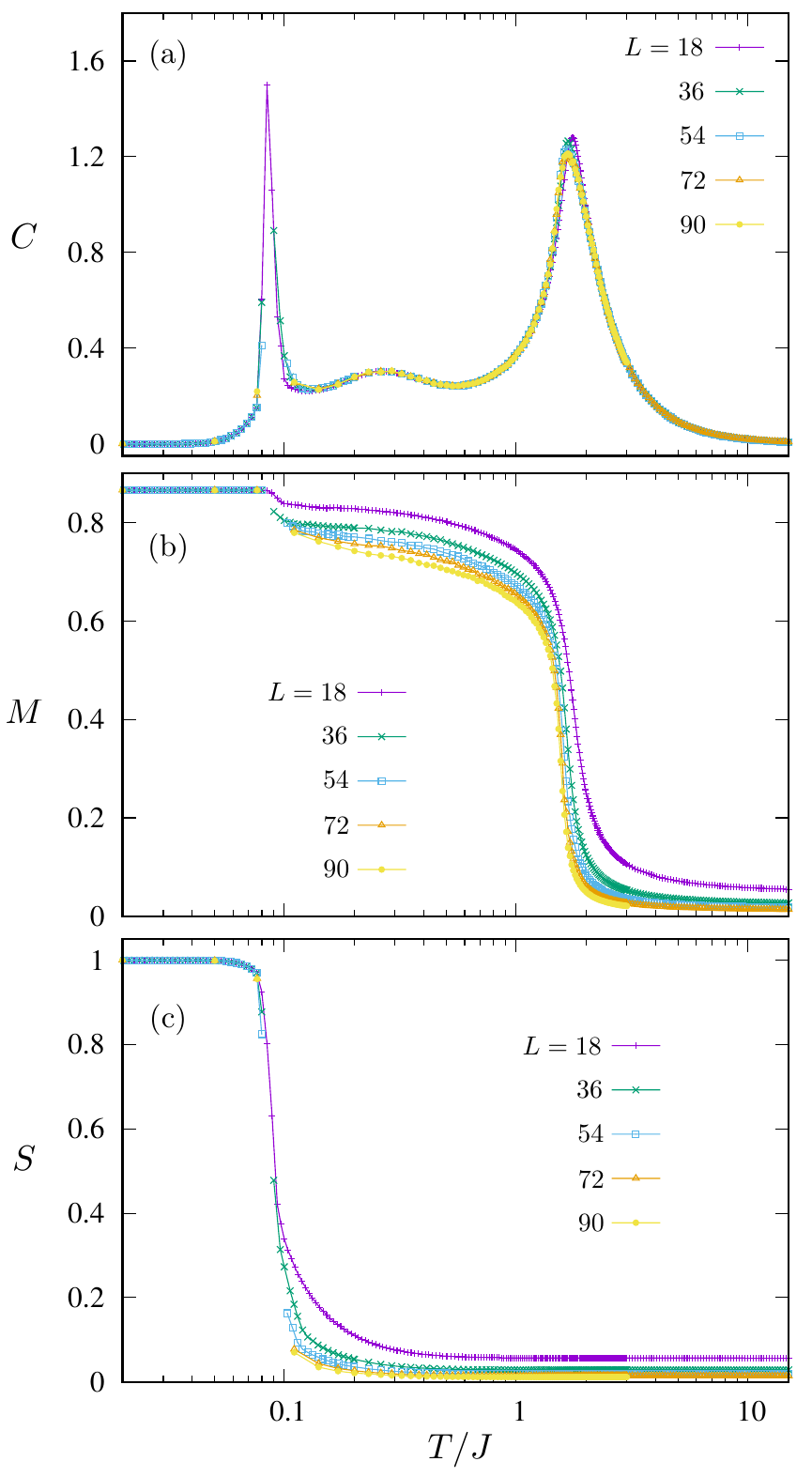}
\caption{\label{f4} (a) Specific Heat $C$, (b) ferromagnetic order $M$, and (c) stripe order parameter as a function of $T/J$ calculated from Monte Carlo simulations for $D/J =0.75$. A pronounced jump in stripe order parameter accompanied by a sharp peak in specific heat indicates a first-order transition into the stripe phase. The two broad peaks at higher temperatures correspond to the two BKT scenario for the breaking of a $Z_6$ symmetry. }
\end{figure}

\begin{figure*}
\includegraphics[width=1.9\columnwidth]{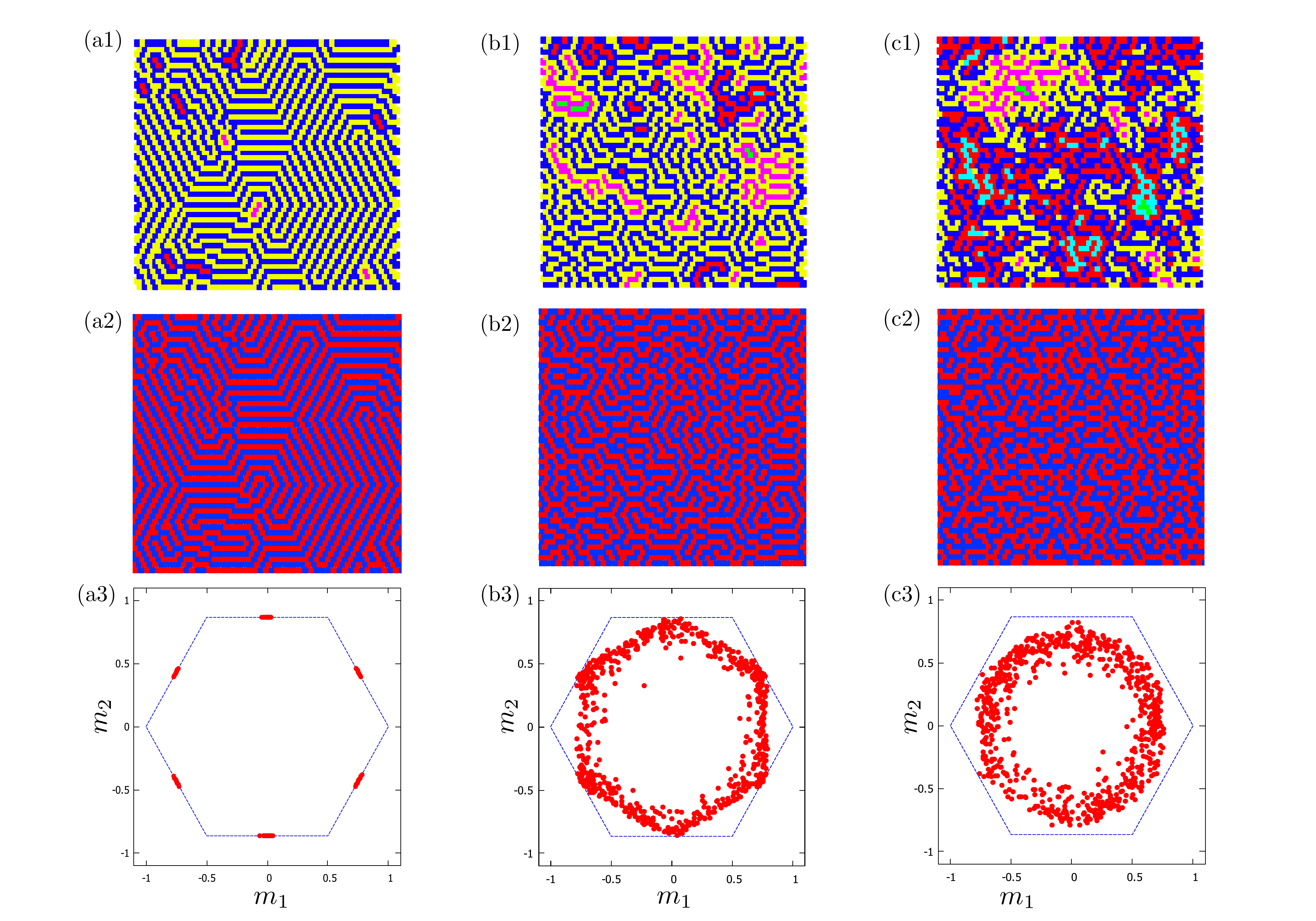}
\caption{\label{f5} (Top) Potts configurations, (middle) Ising configurations, and (bottom) histograms of local ferromagnetic order $\mathbf m = (m_1, m_2)$ for (a) the stripe phase at $T < T_3$, (b) an maze-like hexagonal liquid at $T_3 < T < T_2$, and (c) the critical XY phase at  $T_2 < T < T_1$. As $T \rightarrow 0$, (a) will become perfectly striped and all points in (a3) will be exactly on the midpoints of the edges of the hexagon. The O(2) symmetry in (c3) is consistent with an emergent critical XY phase in the BKT theory.}
\end{figure*}


To explore the interplay between the ferromagnetic ordering and the emergence of maze-like/stripe phases, we next simulate a system with a relatively larger dipolar term $D = 0.75J$. FIG.~\ref{f4} shows the thermodynamic evolution of the system at this value of $D/J$ for 5 different lattice sizes, $L = 18, 36, 54, 72, $ and $90$. In addition to the specific heat and ferromagnetic order parameter, an order parameter $S$ is introduced to characterize the stripe phase. First, we define three local bond parameter associated with a site-$i$: $b_{i, m} = \frac{1}{2} ( \sigma_i \sigma_{i + \hat{\mathbf e}_m} + \sigma_i \sigma_{i - \hat{\mathbf e}_m})$, where $\hat{\mathbf e}_1 = (1, 0)$, $\hat{\mathbf e}_2 = (-\frac{1}{2}, +\frac{\sqrt{3}}{2})$, and $\hat{\mathbf e}_3 = (-\frac{1}{2}, -\frac{\sqrt{3}}{2})$  are unit vectors along the three principal directions of the triangular lattice. And $ i \pm \hat{\mathbf e}_m$ is a short-hand notation for neighboring sites along the $m$-th principal direction. Next, a local doublet vector is introduced to measure the disparity between the three directions
\begin{eqnarray}
	\mathbf f_i = \left( \frac{b_{i, 1} + b_{i, 2} - 2 b_{i, 3}}{\sqrt{6}}, \frac{b_{i, 1} - b_{i, 2}}{\sqrt{2}} \right).
\end{eqnarray}
The stripe order is then defined as
\begin{eqnarray}
	S = \Bigl\langle \Bigl| \frac{1}{N}  \sum_i \mathbf f_i \Bigr| \Bigr\rangle.
\end{eqnarray}
The MC simulation results summarized in FIG.~\ref{f4} find a similar scenario of two BKT transitions, as demonstrated by the two broad peaks in specific heat at higher temperatures $T_1 \approx 1.9$ and $T_2 \approx 0.28$. The ferromagnetic order parameter $M$ shows a similar behavior during this 2 BKT window: a quick rise at the first critical point $T_2$, followed by a relatively smooth increase across $T_2$.  On the other hand, while the stripe order remains very small for this temperature range, a pronounced jump of $S$ takes place at a lower temperature $T_3 \approx 0.08$, indicating the onset of stripe order. This transition into the stripe order is also highlighted by a prominent peak in the specific heat.  Both features suggest a discontinuous transition into the stripe ordered ground state. 

FIG.~\ref{f5} shows snapshots of the various phases below $T_1$ in terms of both Potts (first row) and Ising (second row) variables. The histograms of local ferromagnetic order $\mathbf m$ at the corresponding phases are also shown in the bottom row of FIG.~\ref{f5}. First we consider the critical phase at $T_2 < T < T_1$, corresponding to column (c). Although both Potts and Ising variables are disordered, the histogram again shows an emergent circular symmetry, consistent with a critical XY phase as expected for the intermediate phase in the two BKT transitions scenario.  Interestingly, the Ising configuration seems to already exhibit a maze-like pattern with local stripe ordering, albeit with a rather short coherence length. 

Below the second BKT transition at $T_2$, the Potts variables of the system are expected to exhibit a long-range ferromagnetic order. However, as discussed above, large domains of ferromagnetic Potts states are energetically costly due to the dipolar interaction. Instead, as shown in FIG.~\ref{f5}(b2), the Ising spins in this regime exhibit a maze-like structure similar to the short-range correlated tetragonal liquid on a square lattice~\cite{Booth1995}. In our case, this maze-like phase exhibits the discrete hexagonal symmetry of the underlying triangular lattice, hence closer to the previously reported hexagonal liquid on a honeycomb lattice~\cite{Ruger12}. The short-range stripe-order of the maze-like structure, however, necessarily disrupts domains of a single Potts variable. As a compromise, a long-range ferromagnetic order consisting of two Potts variables $p$ and $ p'$ next to each other on the clock face (i.e. $p - p'  = \pm 1$ mod 6) is developed to accommodate the local stripe order of the Ising variables. Since the Ising variables corresponding to two adjacent Potts states are opposite to each other, such mixed  structures are a compromise of the competing ferromagnetic short-range coupling and the antiferromagnetic dipolar interaction.

The system enters a stripe phase at $T < T_3$. The snapshots in both FIG.~\ref{f5}(a1) and (b1) clearly show domains of straight stripes running parallel to one of the three principal directions of the triangular lattice. On symmetry grounds, the stripe-ordering transition at $T_3$ is similar to the order-disorder transition of the 2D three-state Potts model, which is known to be a continuous transition~\cite{Wu82}. However, instead of the 2D three-state Potts universality class, our simulations found a strong first-order transition into the stripe phase. This is also in contrast to the seemingly continuous stripe-tetragonal or stripe-hexagonal phase transitions~\cite{Booth1995,Ruger12}. 

The origin of the discontinuous transition might be related to a re-entrant behavior of the Potts spins. Consider a perfectly ordered state with stripes running along one of the principal directions. Minimization of the nearest-neighbor ferromagnetic interaction requires a uniform Potts state for individual stripes and adjacent Potts states for neighboring stripes. Importantly, when moving from one stripe to the next, the Potts variable could either increase or decrease by one. The Potts configuration across an array of stripes can then be mapped to the trajectory of a random walker on a clock. The fact that each random step is independent of each other thus indicates the absence of long-range Potts order.

\section{Summary and Outlook}
\label{sec:summary}

To summarize, we have conducted a detailed numerical investigation of a six-state clock model with long-range dipolar interactions, inspired by the ferroelectric ordering in multiferroic hexagonal manganites. At low temperatures, trimerization of local atomic structures results in six distinct but energetically degenerate structural distortions, which are well-described by a six-state clock model. Additionally, the atomic displacements in the trimerized phase generate a local electric polarization, with its sign determined by whether the clock variable is even or odd. These resulting electric dipoles, modeled as emergent Ising degrees of freedom, interact via long-range dipolar couplings.

Through extensive Monte Carlo simulations, we explore the low-temperature phases arising from these competing interactions. As the temperature decreases, the system exhibits two Berezinskii-Kosterlitz-Thouless (BKT) transitions, consistent with the behavior of the standard two-dimensional six-state clock model. The long-range dipolar interactions between the emergent Ising spins drive a first-order transition to a ground state featuring a three-fold degenerate stripe order. Between this first-order transition and the second BKT transition lies an intermediate phase resembling a maze-like hexagonal liquid with short-range stripe correlations. In the case of large dipolar interaction $D \sim J$, this phase also displays an unconventional ferromagnetic order, where adjacent clock variables align to occupy the two distinct stripe types of the mazelike structure. In terms of Potts variables, the system also exhibits an re-entrant behavior when entering the stripe-ordered ground states.

Several open questions remain to be investigated for this interesting system. For example, previous works on dipolar Ising systems show that both the width of the stripe and the transition temperature $T_3$ increase with decreasing dipolar strength. Although our preliminary simulations on the small $D$ systems found a ferromagnetic order consisting of a single Potts state, this result is likely a finite size artifact as the lattice size $L$ could be smaller than the equilibrium stripe width. In the thermodynamic limit, the transition to the stripe order might preempt the second BKT transition and the intermediate ferromagnetic order of Potts spins. As already discussed above, the interplay of local ferromagnetic ordering and stripe formation in the small-$D$ regime might lead to unusual phase ordering dynamics. In general, the highly frustrated nature of the dipolar six-state clock model studied in this work also indicates intriguing nonequilibrium properties and potential glassy behaviors. Finally, a systematic study of the 3D version of this model will be important for a complete modeling of the physics of hexagonal manganites.

\bigskip

\section*{Acknowledgements}

GWC and SZL thank C. D. Batista and Y. Kamiya for collaboration on a related project and useful discussions. The work at University of Virginia was partially supported by the US Department of Energy Basic Energy Sciences under Contract No. DE-SC0020330. The work at LANL (SZL) was carried out under the auspices of the U.S. DOE NNSA under contract No. 89233218CNA000001 through the LDRD Program. The authors also acknowledge the support of Research Computing at the University of Virginia.

\bibliography{reference}
\end{document}